\documentclass[a4paper,fleqn,10pt]{cas-sc}

\usepackage[authoryear,longnamesfirst]{natbib}
\usepackage{amssymb}
\usepackage{amsmath,amsfonts}
\usepackage{algorithmic}
\usepackage{algorithm}
\usepackage{array}
\usepackage[caption=false,font=normalsize,labelfont=sf,textfont=sf]{subfig}
\usepackage{textcomp}
\usepackage{stfloats}
\usepackage{url}
\usepackage{verbatim}
\usepackage{graphicx}
\usepackage{cite}
\usepackage{bigints}
\usepackage{boldline,multirow}
\usepackage{booktabs}
\usepackage{diagbox}
\usepackage{xcolor}
\usepackage [figuresright] {rotating}

\def\tsc#1{\csdef{#1}{\textsc{\lowercase{#1}}\xspace}}
\tsc{WGM}
\tsc{QE}
\tsc{EP}
\tsc{PMS}
\tsc{BEC}
\tsc{DE}

\newcommand{\model}{EvoPath}

\begin{document}
\let\WriteBookmarks\relax
\def\floatpagepagefraction{1}
\def\textpagefraction{.001}

\shorttitle{Evolutionary Meta-path Discovery with LLMs for Complex Heterogeneous Information Networks}

\shortauthors{Liu et~al.}

\title [mode = title]{\model: Evolutionary Meta-path Discovery with Large Language Models for Complex Heterogeneous Information Networks}                      
\tnotemark[1]

\tnotetext[1]{This work was supported in part by the National Natural Science Foundation of China (NSFC, 62206303) and Science and Technology Innovation Program of Hunan Province (Grant 2023RC3009).}

%
\author[inst1]{Shixuan Liu}[orcid=0000-0003-4780-3890]\fnref{fn1,fn2}\ead {liushixuan@nudt.edu.cn}
\author[inst1]{Haoxiang Cheng}\fnref{fn1,fn2}\ead {hx_chenggfkd@nudt.edu.cn}
\author[inst2]{Yunfei Wang}[orcid=0009-0003-5614-9074]\fnref{fn2}
\author[inst3]{Yue He}[orcid=0009-0009-1536-1179]\fnref{fn2}
\author[inst1] {Changjun Fan}\fnref{fn2}\corref{cor2}\ead {fanchangjun@nudt.edu.cn}
\author[inst1]{Zhong Liu}\fnref{fn2}\corref{cor1}

\affiliation[inst1]{organization={Laboratory for Big Data and Decision, College of Systems Engineering, National University of Defense Technology},
            city={Changsha},
            state={Hunan},
            country={China}}

\affiliation[inst2]{organization={National Key Laboratory of Information Systems Engineering, College of Systems Engineering, National University of Defense Technology},
            city={Changsha},
            state={Hunan},
            country={China}}
\affiliation[inst3]{organization={Department of Computer Science and Technology, Tsinghua University},
            city={Beijing},
            country={China}}

\cortext[cor1]{Corresponding author}

\cortext[cor2]{Principal corresponding author}
\fntext [fn1] {Shixuan Liu and Haoxiang Cheng contributed equally as first authors.}
\fntext [fn2] {Declarations of interest: None}
\fntext [fn3] {Digital Object Identifier https://doi.org/10.1016/j.ipm.2024.103920}


\begin{abstract}
Heterogeneous Information Networks (HINs) encapsulate diverse entity and relation types, with meta-paths providing essential meta-level semantics for knowledge reasoning, although their utility is constrained by discovery challenges.
While Large Language Models (LLMs) offer new prospects for meta-path discovery due to their extensive knowledge encoding and efficiency, their adaptation faces challenges such as corpora bias, lexical discrepancies, and hallucination.
This paper pioneers the mitigation of these challenges by presenting \model, an innovative framework that leverages LLMs to efficiently identify high-quality meta-paths.
\model~is carefully designed, with each component aimed at addressing issues that could lead to potential knowledge conflicts.
With a minimal subset of HIN facts, \model~iteratively generates and evolves meta-paths by dynamically replaying meta-paths in the buffer with prioritization based on their scores.
Comprehensive experiments on three large, complex HINs  with hundreds of relations demonstrate that our framework, \model, enables LLMs to generate high-quality meta-paths through effective prompting, confirming its superior performance in HIN reasoning tasks. Further ablation studies validate the effectiveness of each module within the framework.
\end{abstract}







\maketitle

\section{Introduction}

\label{sec:intro}
The impact of Heterogeneous Information Networks (HINs) is increasingly pronounced in diverse domains, providing innovative and structured ways of representing, analyzing, and utilizing typed knowledge bases (KB)~\citep{mitchell2018never}. 
These networks are instrumental in modeling complex systems like social and biological networks~\citep{liu2023hnerec}, knowledge graphs~\citep{Ramaciotti2021Measuring} and recommendation systems~\citep{xun2024higher}, providing richer and more comprehensive insights than traditional homogeneous networks. 
By representing the intricate web of varying types of entities and their diverse connections, HINs support more accurate predictions, in-depth analysis, and personalized recommendations. 
For example, HINs have been vital in mining commercial activities~\citep{xun2024higher}, simulating the structure and dynamic behavior of academic events~\citep{sun2009ranking}, and predicting drug-target interactions~\citep{gonen2012predicting}.

\begin{figure*}[ht]
\centering
\includegraphics[width=1\linewidth]{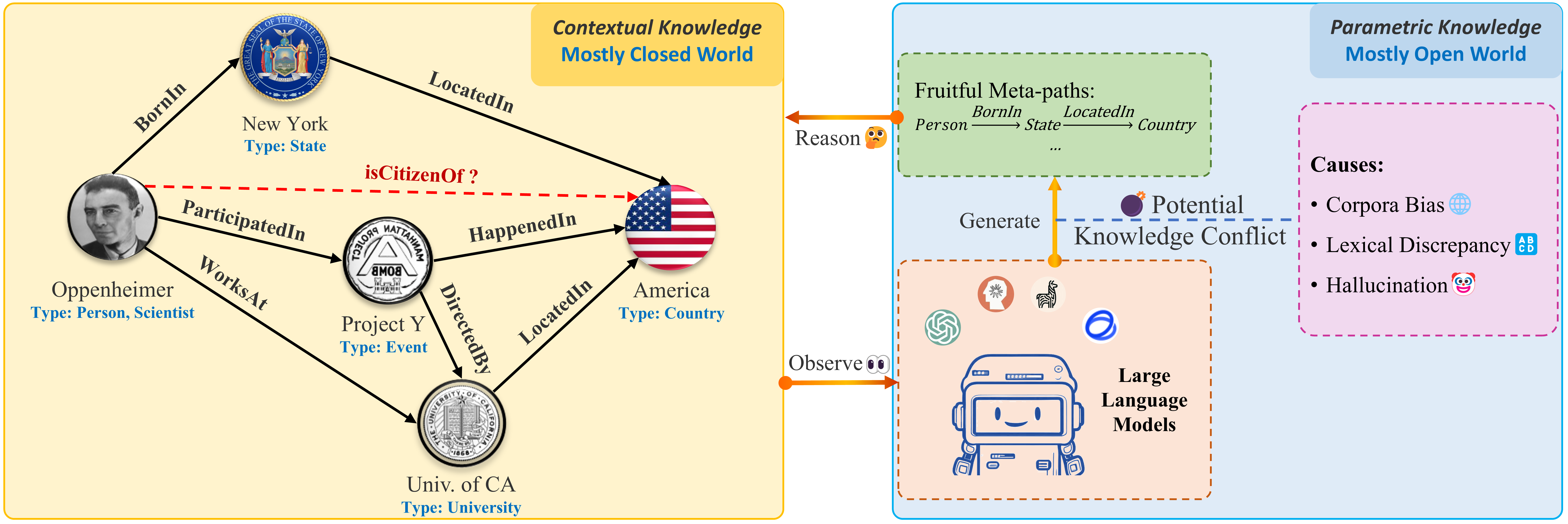}
\caption{While meta-paths offer effective and explainable reasoning, their application is limited by the difficulties inherent in their discovery. Although LLMs present an opportunity for discovering meta-paths, integrating them faces the challenge of potential knowledge conflicts.}\label{fig1}
\end{figure*} 

Meta-path is a high-level abstraction tool in HINs that illuminates the intricate structures within these networks.
Uniquely, a meta-path defines a sequence of entity types and relations that connects two given entities.
For instance, as depicted in Figure ~\ref{fig1}, multiple meta-paths exist between the entity pair \textit{(Oppenheimer, USA)}, including:

\noindent $\text{Person} \xrightarrow{BornIn} \text{State} \xrightarrow{LocatedIn} \text{Country};$

\noindent $\text{Scientist} \xrightarrow{WorksAt} \text{University} \xrightarrow{LocatedIn} \text{Country};$

\noindent $\text{Scientist} \xrightarrow{ParticipatedIn} \text{Event} \xrightarrow{HappenedIn} \text{Country}.$

\noindent These meta-paths effectively support the claim that \textit{Oppenheimer} is a citizen of the \textit{USA}, while also providing explainable insights for other citizenship-related inquiries.
Given their ability to explain complex network structures, meta-paths are crucial for reasoning tasks in HINs under both transductive and inductive scenarios~\citep{sun2011co,kong2012meta}.

Despite their benefits, the usefulness of meta-paths in HIN reasoning is constrained by the challenge associated with their discovery~\citep{liu2023inductive}, which is notably attributed to three factors: the expansive meta-path space, the complexities in assessing meta-path effectiveness, and the neglect of semantic similarity. Specifically, for an HIN with $|T|$ entity types and $|R|$ relations, the potential number of $l$-length candidate meta-paths is $|T| \times (|R| \times |T|)^{l-1}$. Furthermore, effective evaluation of meta-paths is complicated because, being schema-level concepts, their plausibility are assessed through instance-level path observations, a process that is either impractical with simple sampling or overly time-consuming with enumeration~\citep{Zhu2022Effective}. 
Lastly, current methods fail to make use of the semantic similarities that meta-paths may encapsulate in explaining specific relations, leading to undesirable inefficiency.

Recent advancements in Large Language Models (LLMs) have significantly transformed several domains within Artificial Intelligence, enhancing task performance and catalyzing extensive research into their vast repository of implicit knowledge~\citep{achiam2023gpt}. 
Notably, this development could open up new opportunities for meta-path discovery, as LLMs inherently encode a vast amount of knowledge within their parameters through pre-training on extensive text corpora~\citep{pan2023Roadmap}.
The commonsense knowledge within LLMs can be swiftly adapted to reason about facts in KBs, which likely originate from a subset of the training corpora of LLMs~\citep{huang2023adaptive}.
Moreover, LLMs' advanced understanding of semantic similarities and contextual nuances allows for the rapid identification of key rules, including the strategic use of synonyms within meta-paths. 
Ultimately, LLMs could process, analyze, and generate large volumes of text rapidly, thereby facilitating efficient generation of meta-paths from complex HINs characterized by extensive meta-path spaces~\citep{wu2023brief}.

Meanwhile, adopting LLMs for meta-path discovery also presents challenges, as they are not inherently designed for HINs, complicating their direct application.
Most offline KBs, due to the challenges in data acquisition and integration, tend to form a closed world, within which certain knowledge and rules may exhibit biases when compared to the rules perceived by LLMs trained on open-world corpora.
Additionally, while LLMs may generate meta-paths whose semantic meanings align with those of high-quality meta-paths in practice, lexical discrepancies can still occur within the constituent atoms (entity types and relations) of these meta-paths.
Finally, the prevalent issue of hallucination in LLMs also poses a risk of generating invalid or meaningless meta-paths.


In this paper, we introduce \model, a sophisticated framework for LLM-based meta-path discovery. 
To resolve knowledge conflicts, we design an in-context learning (ICL) method tailored for this meta-path discovery. This generation process begins with a pool of random meta-path samples inherent in HINs, guided by predefined criteria for meta-path generation.
To enhance the efficiency of identifying high-quality meta-paths, we draw inspiration from evolutionary algorithms. We prioritize selecting high-quality meta-paths as few-shot samples for LLMs in the ICL process. By leveraging the LLMs' superior language comprehension and generative capabilities, we iteratively enrich the pool with newly generated meta-paths. Low-quality samples in the pool have a minimal likelihood of being selected for further use.
With just a few meta-path examples and their score signals, \model~can iteratively generate numerous high-quality meta-paths for complex HINs.

Main contributions of this paper are summarized as follows:

\begin{enumerate}
    
    \item We exploit the commonsense knowledge of LLMs to generate high-quality meta-paths.

    \item Our LLM-based framework resolves knowledge conflicts between LLMs' parametric knowledge and HINs' contextual knowledge through a novel ICL learning method designed for meta-path discovery and evolutionary techniques.

    \item Extensive experiments conducted on three large, complex HINs demonstrate the superior performance of \model~in HIN reasoning tasks. Ablation studies confirm the effectiveness of each component within \model~and show that \model~is robust to the base LLM model.
\end{enumerate}

The remainder of this paper is structured as follows. Section 2 reviews related work in the field. Definitions necessary for our discussion are introduced in Section 3, followed by a detailed explanation of \model. Section 4 reports on the results of link prediction and knowledge base completion experiments, benchmarking \model~against leading baseline methods across three real-world HINs. Extensive ablation studies are presented in Section 5. Finally, we conclude the paper in Section 6.
\section{Related Work}


\subsection{Meta-path Discovery} 
For HIN reasoning methods based on meta-path, the discovery of relevant meta-paths is a critical prerequisite step.
The most straightforward approach to generating meta-paths involves exhaustive enumeration up to a certain length~\citep{wang2016relsim}. 
To boost efficiency, graph-traversal techniques such as breadth-first search~\citep{kong2012meta} and the $A^{\ast}$ algorithm~\citep{zhu2020effective} have been applied to network schemas for meta-path generation. 
Yet, these methods are limited by the absence of adequate instance-level signals defining the meta-path quality, which hampers the discovery process.
Specifically, \textit{Autopath} employs deep content embedding and continuous Reinforcement Learning (RL) to learn implicit meta-paths, subsequently estimating similarity scores as the empirical probabilities of arriving target entities~\citep{yang2018autopath}.
Some recent works attempt to automate meta-path design with evolutionary search~\citep{han2020genetic} and differentiable structure learning~\citep{ding2021diffmg}. None of the above methods could scale to complex HINs with hundreds of entity types or relations.

To consider meta-path discovery for complex HINs, the majority of approaches follows a two-phase process encompassing the generation followed by summarization of path instances. 
Lao and Cohen advocate for the use of random walks to generate meta-paths within fixed length $l$, incorporating a learnable proximity. However, the choice of $l$ is exceedingly critical to performance and varies greatly across datasets~\citep{lao2010pcrw}. 
\textit{FSPG} employs a greedy method that utilizes user input to identify meta-paths for given entity pairs~\citep{meng2015fspg}.
Wan et al. introduce \textit{MPDRL}, an RL-base multi-hop approach that incorporates type context during walk~\citep{wan2020mpdrl}.
The performance of all above methods may be constrained by their path-finding components due to their reliance on partial observations of path instances.
In particular, \textit{SchemaWalk} is the first to frame meta-path discovery as a Markov Decision Process within the schema graph, with reward signals sourced from the instance graph, enabling the efficient learning of meta-paths with high coverage and confidence ~\citep{liu2023inductive}.
Despite their advancements, all current meta-path discovery methods adhere to the symbolic framework, leaving the potential of leveraging semantic similarity for meta-path discovery unexplored.

\subsection{Knowledge Reasoning with LLMs} 
Considering the strengths of KBs in dynamic, explicit, and structured knowledge representation, various techniques have been developed to integrate LLMs for fact reasoning within KBs~\citep{luo2023Reasonongraph}. 
Despite KBs generally not constituting open-world environments that LLMs are typically trained in, some commonsense rules learned by LLMs in open-world contexts remain relevant to these closed systems~\citep{pan2023Roadmap}.
Currently, LLM-based knowledge reasoning approaches predominantly concentrate on KGs, often overlooking the critical type information in HINs.

Early research efforts focused on embedding structured knowledge from KGs into LLMs either during pre-training or fine-tuning stages, but the inherent explainability of knowledge reasoning and the flexibility of knowledge updates are compromised~\citep{hu2023survey}.
Subsequent studies instead translate relevant structured knowledge from KGs into textual prompts for LLMs, enhancing these prompts with additional information retrieved from KGs~\citep{sun2023think}.
The above methods, aimed primarily at instance-level reasoning, seldom generate rules and lack design for rule mining, which requires LLMs to grasp both the structure of KGs and the semantics of relations to produce meaningful rules.
Specifically, \textit{ChatRule} samples and feeds several relation paths from KGs into LLMs, prompting the generation of meaningful logical rules for reasoning~\citep{luo2023chatrule}.

In complex HINs, the application of LLMs for both instance-level reasoning and meta-path discovery remains unexplored. Our method unlocks opportunities for deploying LLMs for reasoning over complex HINs.
\section{Method}
In this paper, we aim to leverage LLMs to discover high-quality meta-paths. To this end, we begin this section with basic definitions related to HINs, meta-paths, and the key evaluation metrics of coverage and confidence, which assess meta-path plausibility. Following the definitions, we introduce \model, by explaining the rationale behind its component design and presenting each component in depth.

\subsection{Definitions and Notations}

\noindent \textbf{Definition 1 (Heterogeneous Information Network, HIN)} A HIN $\mathcal{H}$ is defined as a directed graph $\mathcal{G}=(V, E, \tau, \phi)$, where $V$ is the set of entities, $E\subseteq V\times V$ represents the edges connecting these entities. The function $\tau:V\rightarrow T$ assigns entity types to an entity from the type taxonomy $T$, and $\phi:E\rightarrow R$ maps edges to relations in the set $R$. Summarizing entities and edges in $G$ into types and relations, we obtain the schema graph $\mathcal{G}_S=(T,R)$.

\noindent\textbf{Definition 2 (Meta-path)} A meta-path $M$ of length $l$ is a path on the schema graph $\mathcal{G}_S$, defined as $M=t_1\xrightarrow{r_1}t_2\xrightarrow{r_2}\cdots\xrightarrow{r_{l-1}}t_l$, with $t_i\in T$ for an entity type and $r_i\in R$ for a relation. 
In this paper, we refer to entity types and relations in meta-paths as atoms.
A path $P=v_1\xrightarrow{r_1}v_2\xrightarrow{r_2}\cdots\xrightarrow{r_{l-1}}v_l$ is a meta-path instance of $M$ if $\forall i\in \{1,\cdots,l\}, t_i\in \tau(v_i)$ and $\forall i\in \{1,\cdots,l-1\}, e_i\in \phi(v_i, v_{i+1})$. 
In this case, the entity pair $(v_1, v_l)$ is connected by a path instance of $M$, represented as $\mathbb I_M(v_1, v_l)$.

Meta-path instances are commonly employed to evaluate the plausibility of meta-paths. In association rule mining, coverage and confidence stand out as crucial metrics for rule evaluation, each providing unique insights~\citep{agrawal1993mining}. Coverage measures the applicability of a rule, whereas confidence evaluates its reliability.

\noindent\textbf{Definition 3 (Coverage)} The coverage of a meta-path $M$ for a given relation $r_q$ quantifies how frequently $M$ appears within $r_q$-related entity pairs.
It is calculated as the ratio of entity pairs connected by both $r_q$ and a path instance of $M$, to the number of entity pairs connected by $r_q$,

\begin{equation}
Cov_{M\Rightarrow r_q}^\mathcal{H}:=\frac{\#(v_i,v_j):\mathbb I_M(v_i,v_j) \wedge r_q\in \phi(v_i, v_j)}{\#(v_i,v_j):r_q\in \phi(v_i, v_j)}, v_i,v_j\in \mathcal{H}
\label{equ_cover}
\end{equation}

\noindent\textbf{Definition 4 (Confidence)} The confidence of a meta-path $M$ in predicting the relation $r_q$ given facts in $\mathcal{H}$ is defined as the ratio of entity pairs connected by both $r_q$ and a path instance of $M$, to all entity pairs connected by any path instance of $M$,

\begin{equation}
Conf_{M\Rightarrow r_q}^\mathcal{H}:=\frac{\#(v_i,v_j):\mathbb I_M(v_i,v_j) \wedge r_q\in \phi(v_i, v_j)}{\#(v_i,v_j):\mathbb I_M(v_i,v_j)}, v_i,v_j\in \mathcal{H}
\label{equ_conf}
\end{equation}

Considering the entity pairs linked by a relation, coverage measures the frequency of these pairs satisfying a meta-path, while confidence assesses the accuracy of a meta-path’s representation of the relation. Analyzing these metrics enables the identification of rules with higher relevance and utility.

\begin{figure*}[ht]
\centering
\includegraphics[width=1\linewidth]{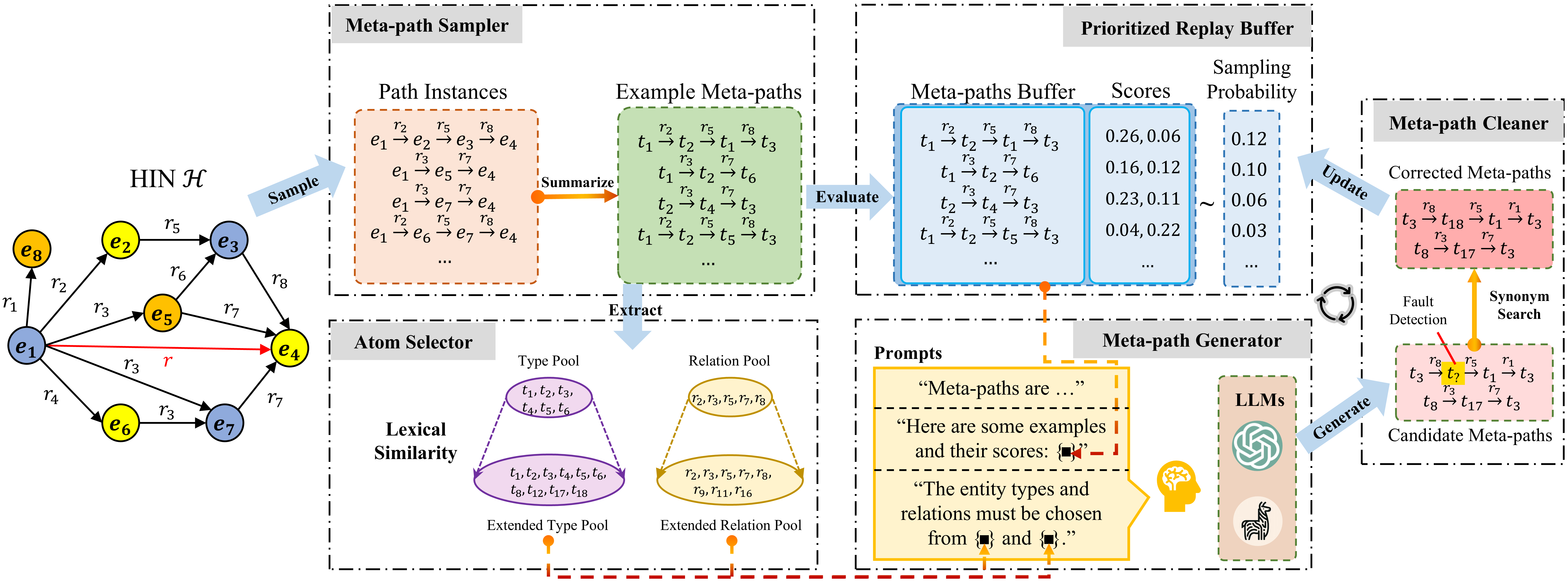}
\caption{Given a HIN, the meta-path sampler initially generates meta-path examples from path instances sampled via random walks, which are then processed by the atom selector and prioritized replay buffer.
The atom selector extracts the taxonomy of entity type and relation from example meta-paths and expands them using lexical similarity to construct candidate atoms. Meanwhile, the prioritized replay buffer calculates plausibility scores for meta-path examples to determine their sampling probabilities.
Subsequently, by integrating the sampled paths and candidate atoms into prompts, the meta-path generator establishes meta-paths with LLMs. 
Ultimately, the meta-path cleaner identifies and corrects errors, considering synonyms where possible, before plugging the corrected meta-paths into the buffer.
A cyclical evolutionary process encompassing the replay buffer, meta-path generator, and cleaner is then initiated, progressively refining the meta-paths.
}\label{fig:model}
\end{figure*} 

\subsection{\model~Framework}
\label{sec:framework}
This subsection details our proposed framework, comprising five key components: 1) a meta-path sampler for the efficient generation of example meta-paths, crucial for enabling effective in-context learning (ICL) to address corpus bias and hallucination; 2) an atom selector to ensure valid meta-path generation by determining candidate taxonomies for entity types and relations; 3) a prioritized replay buffer for sampling meta-paths based on plausibility scores, introducing a novel ICL technique; 4) a meta-path generator that employs large language models (LLMs) to produce meta-paths, prompted with sampled meta-paths and lexical constraints, further addressing lexical discrepancies; and 5) a meta-path cleaner for error correction in generated meta-paths.
The overall workflow is illustrated in Figure~\ref{fig:model} and each component is elaborated as below.

\noindent\textbf{Meta-path Sampler.} To enhance LLMs' comprehension of HIN structures for improved meta-path discovery, we initially gather path instances within the HIN, representing meta-paths at the entity level~\citep{cheng2023NCRL}.
To efficiently sample these path instances, we adopt a procedure based on random walks~\citep{spitzer2013principles}.
Specifically, we initiate by sampling a batch of facts $\{(v_h, r, v_t)\}$ for the $r$ requiring reasoning, then simulate a fixed-length $l$ random walk from source entity $v_h$, with the node $e_i$ at each step $i$ generated based on following distribution:

\begin{equation}
\label{equ:walk}
P(v_i|v_{i-1}) = 
\begin{cases}
      1/|N(v_{i-1})|, & \exists r: (v_i, r, v_{i-1}) \in \mathcal{H} \\
      0, & \text{otherwise}
    \end{cases} 
\end{equation}
where $N (\cdot)$ denotes the neighborhood. 
To boost efficiency, we add a step to the random walk process, checking for the existence of the relation $r$ between the starting node $v_s$ and the current node $v_i$, and if such edge exists, we record a path instance.
With above process, sufficient evidence explaining a relation is collected, poised for summarization into meta-paths.
Considering entities may have multiple types, we summarize these path instances by identifying the Lowest Common Ancestor in a type directed acyclic graph, thereby generating diverse meta-paths.

\noindent\textbf{Atom Selector.} To address the lexical discrepancy between sampled and LLM-generated meta-paths (e.g., LLM could use \textit{Citizenship} instead of \textit{isCitizenOf}), we confine LLM output to a predefined taxonomy of valid relations and entity types. 
Furthermore, the complexity of some HINs (e.g., $R=827$ and $T=756$ for NELL) precludes the direct use of the complete sets of $T$ and $R$ as taxonomy, otherwise presenting substantial challenges for LLMs in combing these taxonomies into meaningful meta-paths.

Since most relations and entity types are redundant for explaining a relation due to their logical distance or conceptual irrelevance (e.g., \textit{hasCurrency} is irrelevant for explaining \textit{isMarriedTo}, or \textit{hasAcademicAdvisor} for \textit{playsFor}), we require the atom selector to only identify relevant taxonomy based on evidence from the HIN. 
Therefore, we initially extract relations and entity types present in sampled meta-paths. 
Besides, considering that sampled meta-paths provide only a partial observation and might result in an overly narrowed taxonomy space, potentially impairing meta-path generation, we expand the extracted taxonomy by conducting semantic similarity searches for each term against $T$ or $R$. We opt for the efficient Gestalt pattern matching for this goal~\citep{ratcliff1988pattern}.

\noindent\textbf{Prioritized Replay Buffer.} In multiple reasoning tasks, providing LLMs with high-quality examples, as opposed to random ones, typically results in significantly improved results~\citep{huang2023boosting}.
While current prompt retrieval methods enhance task performance by selecting high-quality few-shot examples through semantic-based heuristics or supervised retrieval models, they are not specifically designed for meta-path discovery.
To address this gap, we develop a novel mechanism for selecting high-quality few-shot meta-paths based on their plausibility scores, which serve as indicators of their quality. 
To mitigate the risk of over-emphasis on a limited dataset due to greedy prioritization, we employ a stochastic prioritization method. This approach maintains a monotonically increasing sampling probability in accordance with the meta-path priority while ensuring that every path—even those with the lowest plausibility scores—has a non-zero chance of being selected. This balanced methodology enhances the quality and diversity of the selected examples, thereby optimizing the performance of meta-path discovery tasks.

We define the sampling probability of meta-path $j$ as $P(j) = p_j / \sum_{k=1}^{|\mathcal{B}|}p_k$, where $p_j$ is the priority of meta-path $j$, and $|\mathcal{B}|$ is the size of the buffer.
The assignment of priority $p_j$ can be implemented either directly or indirectly: directly as $p_j = score(j)$, or indirectly through a rank-based approach where $p_j=1/rank(j)$, with $rank(j)$ determined by sorting meta-paths according to their respective $score(j)$.
The score $score(j)$ may be derived from various criteria such as coverage, confidence, or a combination of both. 
Empirically, we find that rank-based prioritization using the combined score yields the highest-quality meta-paths. To determine the most effective method for priority assignment and scoring, we conduct comprehensive ablation studies, detailed in Section~\ref{sec:ablation}.

\noindent\textbf{Meta-path Generator with LLM.} 
We carefully designed prompts to incorporate both structural and semantic information through ICL, thereby optimizing the utilization of LLMs for meta-path discovery. 
Our approach involved structuring the prompts into a coherent narrative, divided into three distinct parts.
Initially, we provide a background overview, familiarizing the LLM with the nature of HINs and the definition of meta-paths. 
Subsequently, we derive few-shot examples with meta-paths sampled using the prioritization technique described earlier. This includes presenting concrete instances along with their respective scores to facilitate the LLM’s understanding.
To enhance comprehension of the semantics embedded within the meta-paths, we translate them into natural language descriptions.
Lastly, we define specific generation constraints, such as the maximum allowable length of meta-paths and the permissible types of entities and relations, to ensure the generation of valid and relevant meta-paths. 
Based on ICL, our method offers two key benefits: it simplifies the integration of contextual knowledge from HINs into LLMs by prioritizing demonstrations and providing requirements, and it eliminates the need for extensive fine-tuning, significantly reducing computational costs.
The detailed prompt structure is available in Appendix~\ref{sec:appendix_prompt}.

\noindent\textbf{Meta-path Cleaner.} Despite the inclusion of few-shot examples and the application of constraints, the LLM might still generate invalid meta-paths due to hallucination. 
To address this issue, we employ a remedial step involving a meta-path cleaner, which identifies and corrects errors in the generated meta-paths, utilizing synonym searches where applicable.
Furthermore, any meta-path sequence is deemed incorrect and discarded if, at any step $i$, $t_i\xrightarrow{r_i}t_{i+1}$ lacks corresponding facts in the HIN.
The validated and corrected meta-paths are subsequently integrated into the replay buffer, serving as potential samples for the next cycle of meta-path generation.
\section{Experiment}

\subsection{Experimental settings}
\noindent\textbf{Tasks.} To evaluate the performance and efficiency of \model, we conducted experiments of knowledge base completion (KBC) and link prediction utilizing the meta-paths generated by our model.
The efficiency of \model~allowed for the derivation of rules for various relations within a reasonable time, enabling effective KBC.
In KBC tasks, our objective is to discover the target entity $v_t$ given a query in the form $(v_s, r_q, ?)$.
Furthermore, in line with the HIN literature where model effectiveness is commonly assessed on a per-relation basis~\citep{meng2015fspg,wan2020mpdrl}, we showcase \model's effectiveness through link prediction tasks, distinguishing between positive and negative facts. 
Finally, we perform inductive reasoning experiments on entities fully unseen in the train graph used for meta-path generation.

\noindent\textbf{Datasets} We conducted experiments across three complex real-world HINs: YAGO26K-906~\citep{suchanek2007yago}, Dbpedia~\citep{auer2007dbpedia}, and NELL~\citep{mitchell2018never}, each characterized by a rich diversity of entity types and relations. YAGO26K-906 and Dbpedia were selected for KBC tasks. For link prediction, we evaluated three relations each in YAGO26K-906 and NELL: \{\textit{isCitizenOf}, \textit{DiedIn}, \textit{GraduatedFrom}\} for YAGO26K-906, and \{\textit{WorksFor}, \textit{CompetesWith}, \textit{PlaysAgainst}\} for NELL. The statistics of the datasets are presented in Table~\ref{table_dataset} and detailed in Appendix~\ref{sec:appendix_data}.
\begin{table}[h!]
    \centering
       \caption{Statistics of real-world datasets}
       \renewcommand\arraystretch{1.2}
\resizebox{0.75\linewidth}{!}{%
 \begin{tabular}{c|cccc}
    \toprule[1.5pt]
        Dataset &\#Entity &\#Entity Types &\#Relation &\#Facts\\
     \hline
        \textbf{Yago26K-906} & 26,078 & 906 & 34 & 390,738  \\
        \textbf{Dbpedia} &  111,762 &174 & 305 & 863,643\\
      \textbf{NELL} &  49,869 &756 & 827 & 296,013\\
       \bottomrule[1.5pt]
   \end{tabular}
   }
   \label{table_dataset}
\end{table}
\begin{table}[htb]
\centering
\caption{Categories of baseline methods}
\resizebox{0.6\linewidth}{!}{
\begin{tabular}{c|c|c|c}
\toprule[1.5pt]
Names & Meta-path-based   & Embedding-based  & Rule-based                               \\ \hline
\textbf{MPDRL}                  &     \checkmark    &        &                              \\ 
\textbf{PCRW}                   &     \checkmark    &        &                           \\ 
\textbf{Autopath}               &        \checkmark &        &             \\ 
\textbf{Metapath2Vec}           &    \checkmark     &  \checkmark      &                    \\ 
\textbf{HIN2Vec}                &   \checkmark      &   \checkmark     &           \\ 
\textbf{HGMAE}               &        \checkmark &   \checkmark     &             \\ 
\textbf{TransE}                 &         &     \checkmark   &          \\ 
\textbf{DistMult}                 &         &     \checkmark   &          \\ 
\textbf{ComplEx}                 &         &     \checkmark   &          \\ 
\textbf{RotatE}                 &         & \checkmark       &                  \\ 
\textbf{RNNLogic}                 &         &        &   \checkmark       \\ 
\textbf{MINERVA}                &         &        &  \checkmark               \\
\textbf{MLN4KB}                &         &        &  \checkmark               \\
\bottomrule[1.5pt]

\end{tabular}
}
\label{tab:baseline}
\end{table}

\noindent\textbf{Baselines.} We benchmark \model~against thirteen leading reasoning methods for HINs, spanning meta-path-based, embedding-based, and rule-based strategies, namely MPDRL~\citep{wan2020mpdrl}, PCRW~\citep{lao2010pcrw}, Autopath~\citep{yang2018autopath}, Metapath2Vec~\citep{dong2017metapath2vec}, HIN2Vec~\citep{fu2017hin2vec}, HGMAE~\citep{tian2023heterogeneous}, TransE~\citep{bordes2013transe}, DistMult~\citep{yang2014Distmult}, ComplEx~\citep{trouillon2016complex}, RotatE~\citep{sun2019rotate}, RNNLogic~\citep{qu2020rnnlogic}, MINERVA~\citep{das2018minerva} and MLN4KB~\citep{fang2023mln4kb}. The categorization of these methodologies is presented in Table~\ref{tab:baseline}, with some methods belonging to multiple categories.
Meta-path-based approaches leverage meta-paths to facilitate reasoning processes, whereas rule-based methods primarily employ logical rules for reasoning, explicitly excluding the type information inherent in meta-paths. On the other hand, embedding-based methods focus on generating embeddings by considering a predefined distance metric.
We run the code provided in their papers. Detailed descriptions of these methods are available in Appendix~\ref{sec:appendix_baseline}.

\noindent \textbf{Dataset Preparation.} 
In the KBC task, we adopt a 9:1 ratio for the training/testing split over facts. Throughout the meta-path generation process, we exclude facts present in the test sets from the instance graph to estimate the plausibility scores. 
For Dbpedia, consisting of 305 relations, we test on a subset of these relations, shown in Appendix~\ref{sec:appendix_data}.

In the link prediction task for a given relation $r_q$, we assess whether each $r_q$-connected entity pair is linked by an instance path of length $l-1$, where $l$ denotes the predefined maximum meta-path length, excluding direct connections through $r_q$.
Pairs not strictly meeting this criterion are subsequently excluded.
Contrary to the approach adopted in MPDRL~\citep{wan2020mpdrl}, we remove the limit on search attempts, allowing for a wider inclusion of compliant pairs for fairer comparison.
The resulting pairs are segmented into training and testing sets at an 8:2 ratio. Test set is omitted from the instance graph to assess coverage and confidence of meta-paths. Following standard practice, we generate negative pairs by substituting the target entity in instance graph samples with a fake entity of the same type. This method is applied to generate all negative samples across the six relations in YAGO26K-906 and NELL, with a positive-to-negative ratio of 2:1.

In our entity-level inductive experiment, we randomly select 40\% of the positive test set and gradually remove different percentages ($0\%$, $20\%$, $50\%$ and $100\%$) of nodes present in these pairs from the instance graph, followed by standard meta-path generation and link prediction processes.

\noindent \textbf{Configurations for \model.} We conduct experiments on a server with a 96-core CPU, an 800 GB memory and four 40GB A100 GPUs. 
We utilize LLama-2-7B as our base LLM to generate meta-paths, employing a rank-based prioritization that based on the sum of confidence and coverage.
Each time, 30 sampled meta-paths and their scores are used for few-shot examples. The maximum meta-path length $l$ is 3.

\noindent \textbf{Metrics.} 
For the KBC task, with the query $(v_h, r_q, ?)$, we rank potential tail entities for $v_h$ based on aggregated (max-pooled) confidence scores of mined meta-paths. 
Entities unreachable via any meta-path from $v_h$ receive an infinite rank. We assess KBC performance using standard metrics: Hits@1, 3, 10, and mean reciprocal rank (MRR). 
In link prediction, we utilize two metrics: the area under the receiver operating characteristic curve (ROC-AUC) and average precision (AP).
All above results are averaged over five independent runs.
\begin{table*}[htb]
    \centering
 \renewcommand\arraystretch{1.4}
       \caption{KBC results on YAGO26K-906 and Dbpedia, averaged over 5 runs. The best/second best metrics are bolded/underlined.}
\resizebox{\linewidth}{!}{%
 \begin{tabular}{cc|ccccccccc}
 \toprule[1.5pt]
 &&\model& TransE & DistMult & ComplEx & RotatE & MINERVA & RNNLogic & AutoPath & MLN4KB
 \\
 \hline
 \multirow{4}{*}{YAGO}& Hits@1 & 0.141 & 0.092 & 0.039 & 0.069& \pmb{0.180} & 0.133&\underline{0.161}
 & 0.118  & 0.127 \\
 & Hits@3 & 0.250 & 0.225 & 0.052 & 0.123& \underline{0.260} & 0.214&\pmb{0.263} & 0.226 & 0.234 
 \\
 & Hits@10 & \pmb{0.397} & 0.316 & 0.118 & 0.174& 0.347 & 0.310&\underline{0.364} & 0.334 & 0.342
 \\
 & MRR & \underline{0.229} & 0.176 & 0.059 & 0.106& \pmb{0.237} & 0.192&0.225 & 0.207 & 0.213
 \\
  \hline
 \multirow{4}{*}{Dbpedia}& Hits@1 & 0.674 & 0.538 & 0.359 & 0.626& \pmb{0.753} & \underline{0.684}&0.658 &0.634 & 0.669
 \\
 & Hits@3 & \pmb{0.839} & 0.809 & 0.481 & 0.728& \underline{0.813} & 0.786&0.731 & 0.696 & 0.722
 \\
 & Hits@10 & \pmb{0.899} & 0.858 & 0.555 & 0.769& 0.841 & \underline{0.863}&0.773 & 0.796 & 0.785
 \\
 & MRR & \underline{0.765} & 0.675 & 0.431 & 0.682& \pmb{0.791} & 0.745&0.708 & 0.698 & 0.710
 \\
 \bottomrule[1.5pt]
\end{tabular}
}
\label{table_multi_transductive}
 \end{table*}

\subsection{Transductive Experiment Results}
\subsubsection{Knowledge Base Completion Results.}

\begin{table}[htb]
    \centering
    \renewcommand\arraystretch{1.35}
       \caption{Link prediction results for YAGO26K-906 and NELL, averaged over 5 runs. The bold/underlined results indicate the best/second best performances.}
\resizebox{1\linewidth}{!}{%
\begin{tabular}{ccc>
 {\centering}m{0.06\textwidth}>{\centering}m{0.06\textwidth}>
 {\centering}m{0.06\textwidth}>{\centering}m{0.06\textwidth}>{\centering}m{0.10\textwidth}>{\centering}m{0.06\textwidth}>{\centering}m{0.06\textwidth}>{\centering}m{0.06\textwidth}>{\centering}m{0.05\textwidth}c}
 
    \toprule[1.5pt]
   &&&\model & MPDRL & PCRW
   & Autopath & Metapath2Vec & HIN2Vec & HGMAE & RotatE & TransE &MINERVA 
     \\
     \hline
      \multirow{6}{*}{YAGO} & \multirow{2}{*}{isCitizenOf}&ROC-AUC&$\pmb{0.949}$&$0.781$&$0.584$&$0.757$&$0.652$&$0.800$&0.808&$0.778$&$0.590$& \underline{$0.828$} 
       \\
      
 & &AP&$\pmb{0.968}$&$0.796$&$0.706$&$0.724$&$0.781$&$0.837$&0.823 &$0.830$&$0.810$& \underline{$0.840$}
      \\
           
     & \multirow{2}{*}{DiedIn}&ROC-AUC&$\pmb{0.901}$&$0.710$&$0.645$&$0.723$&$0.661$&$0.785$&0.794&\underline{$0.864$}&$0.622$&$0.632$
       \\
       & &AP&$\pmb{0.942}$&$0.679$&$0.686$&$0.787$&$0.830$&$0.877$&0.838&\underline{$0.909$}&$0.745$&$0.786$
      \\
                
     & \multirow{2}{*}{GraduatedFrom}&ROC-AUC&$\pmb{0.831}$&$0.664$&$0.586$&$0.724$&$0.661$&$0.803$&0.792&\underline{$0.817$}&$0.662$&$0.609$
       \\
       &&AP&$\pmb{0.899}$&$0.743$&$0.664$&$0.718$&$0.783$&$0.842$&0.829&\underline{$0.847$}&$0.750$&$0.718$

     \\
     \hline
     \multirow{6}{*}{NELL} & \multirow{2}{*}{WorksFor}&ROC-AUC&$\pmb{0.897}$&$0.759$&$0.646$&$0.703$&$0.613$&$0.790$&0.794&\underline{$0.868$}&$0.637$&$0.767$
       \\
       &&AP&$\pmb{0.935}$&$0.871$&$0.735$&$0.778$&$0.819$&$0.870$&0.872&\underline{$0.911$}&$0.719$&$0.802$
      \\
                   
      &\multirow{2}{*}{PlaysAgainst}&ROC-AUC&\underline{$0.957$}&$0.774$&$0.567$&$0.544$&$0.761$&$0.783$&0.843&$\pmb{0.966}$&$0.845$&$0.541$
       \\
       &&AP&\underline{$0.973$}&$0.823$&$0.745$&$0.691$&$0.904$&$0.917$&0.890&$\pmb{0.993}$&$0.928$&$0.683$
      \\
                   
      &\multirow{2}{*}{CompetesWith}&ROC-AUC&\underline{$0.773$}&$0.589$&$0.547$&$0.567$&$0.600$&$0.730$&0.715&$\pmb{0.870}$&$0.771$&$0.585$
       \\
       &&AP&$0.855$&$0.695$&$0.699$&$0.685$&$0.735$&$0.867$&0.834&$\pmb{0.939}$&\underline{$0.908$}&$0.702$
      \\
       \bottomrule[1.5pt]
   \end{tabular}
   }
   \label{table_link_result}
\end{table}
Table~\ref{table_multi_transductive} presents the KBC performance of \model~on YAGO26K-906 and Dbpedia datasets. 
\model~exceeds the capabilities of three embedding-based methods—TransE, DistMult and ComplEx—and notably surpasses the path-based MINERVA and meta-path-based Autopath. On YAGO26K-906, it leads in the Hits@10 metric and nearly matches the best performing embedding-based RotatE in terms of MRR. 
For Dbpedia, \model~secures the highest Hits@3 and Hits@10 scores, along with the second-best MRR score, demonstrating robust competition against cutting-edge models in KB reasoning.
While RotatE marginally outperforms \model~on Dbpedia, its embedding-based nature makes it susceptible to unseen entities, as discussed in Section~\ref{sec:inductive}.

\subsubsection{Link Prediction Results.}
Table~\ref{table_link_result} details the link prediction results for six selected relations. \model~exceeds performance across all YAGO26K-906 relations and the \textit{WorksFor} relation in NELL, closely rivaling RotatE in the \textit{PlayAgainst} relation. Despite not outperforming some embedding baselines in the \textit{CompetesWith} relation, \model~still achieves a relatively high performance.

Among meta-path-based baselines, HIN2Vec stands out by producing superior embeddings through utilizing meta-paths for multiple relations, outperforming Metapath2Vec. The performance of HGMAE is comparable to that of HIN2Vec, demonstrating its capability to effectively capture complex graph structures. MPDRL achieves satisfactory outcomes through its RL strategy for path exploration, while Autopath also offers decent performance, albeit constrained by its discovery of a limited number of meta-paths. PCRW ranks as the least effective model due to its reliance on randomness. The path-based MINERVA underperforms in certain relations due to its inability to utilize entity type information.
Embedding approaches for knowledge graphs, particularly RotatE, are competitive in NELL's relations. We discovered that these relations often encompass vast meta-path spaces, posing significant challenges to most meta-path-based methods.

\subsection{Inductive Experiments Results}
\label{sec:inductive}
\begin{figure*}[htb]
\centering
\includegraphics[width=\linewidth]{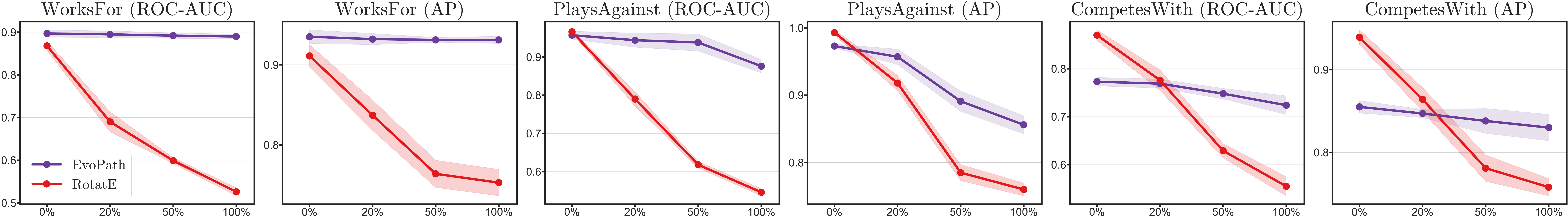}
\caption{Inductive link prediction results for \model~(Purple) and RotatE (Red). The horizontal axes denote node removal rate, and the shaded areas represent confidence intervals across five runs.
}\label{fig_inductive}
\end{figure*} 

It is important to note that directly comparing meta-path-based methods with embedding methods in transductive completion tasks is not entirely fair. Meta-path-based methods excel in inductive scenarios by reasoning over unseen entities, a capability that embedding methods lack.

In the inductive experiment, we focus on the NELL dataset, where our method did not excel. The results are shown in Figure~\ref{fig_inductive}. We observe a sharp decline in RotatE's performance as the removal rate increases, in contrast, \model~displays only minimal impact. At a $50\%$ removal rate, \model~clearly outperforms in all chosen relations. Remarkably, even at the highest removal rate, \model's performance in the inductive setting surpasses that of other baselines in transductive settings.

\begin{table}[htb]
\centering
\caption{Example meta-paths found by \model }\label{tab_example_metapath}
\renewcommand\arraystretch{1}
\resizebox{0.9\linewidth}{!}{%
\begin{tabular}{c|c|c|c}
\toprule[1.5pt]
\textbf{Relations}& \textbf{Meta-path}&\textbf{Conf.}&\textbf{Cov.}
\\ \midrule[1.1pt]
\multirow{5}{*}{\textit{isCitizenOf}}   
& $\mbox{Person} \xrightarrow{WorksAt}\mbox{University}\xrightarrow{LivesIn}\mbox{Country}$   &0.366&0.138                     \\

& $ \mbox{Person} \xrightarrow{WasBornIn}\mbox{Commune}\xrightarrow{hasCapital^{-1}}\mbox{Country}$ &0.238 &0.141                       \\
             
& $\mbox{Person}\xrightarrow{WasBornIn}\mbox{District}\xrightarrow{LocatedIn}\mbox{Country}$      &  0.295 & 0.083                   \\

&  $\mbox{Person} \xrightarrow{GraduatedFrom}\mbox{University}\xrightarrow{LocatedIn}\mbox{Country}$      &  0.246 & 0.134                   \\

&  $\mbox{Scientist} \xrightarrow{DiedIn}\mbox{Location}$&0.438 &0.166     \\

\hline
\multirow{5}{*}{\textit{GraduatedFrom}}   
&$\mbox{Person} \xrightarrow{WorksAt}\mbox{University}$      &  0.218 & 0.125                   \\
& $\mbox{Person} \xrightarrow{isCitizenOf}\mbox{Country}\xrightarrow{LocatedIn^{-1}}\mbox{University}$      &  0.169 & 0.008                   \\
& $\mbox{Person} \xrightarrow{hasAcademicAdvisor}\mbox{Scientist}\xrightarrow{WorksAt}\mbox{University}$      &  0.029 & 0.307                   \\
& $\mbox{Officeholder} \xrightarrow{diedIn}\mbox{Town}\xrightarrow{LocatedIn^{-1}}\mbox{University}$      &  0.293 & 0.003                   \\
&$\mbox{Economist} \xrightarrow{BornIn}\mbox{Administrative}\xrightarrow{LocatedIn^{-1}}\mbox{University}$      &  0.333 & 0.002                  \\

\hline
\multirow{5}{*}{\textit{WorksFor}}   

& $\mbox{Person} \xrightarrow{ControlledBy}\mbox{Company}$      &  0.423 & 0.647                   \\
& $\mbox{Chef} \xrightarrow{Writesfor}\mbox{Company}$          &0.981&0.002        \\

&$\mbox{CEO} \xrightarrow{Leads}\mbox{Company}$      &  0.947 & 0.598                   \\
& $\mbox{Journalist} \xrightarrow{WritesFor}\mbox{Company}$      &  0.434 & 0.309                   \\
&$\mbox{Writer} \xrightarrow{Leads}\mbox{Book}$ &0.943&0.018                 \\

\hline
\multirow{5}{*}{\textit{PlayAgainst}}   
&$\mbox{Team} \xrightarrow{SubpartOf}\mbox{League}\xrightarrow{LeagueTeam}\mbox{Team}$      &  0.625 & 0.280                   \\
& $\mbox{Team} \xrightarrow{PlaysIn}\mbox{League}\xrightarrow{LeagueTeam}\mbox{Team}$      &  0.589 & 0.251                   \\
&$\mbox{Team}\xrightarrow{SubpartOf}\mbox{League}\xrightarrow{SubpartOf}\mbox{Team}$&0.285&0.395                    \\
& $\mbox{Team} \xrightarrow{Won}Game\xrightarrow{CoachedBy}\mbox{Coach}\xrightarrow{BelongsTo}\mbox{Team}$ &0.323&0.108                      \\
&$\mbox{Coach} \xrightarrow{Playsport}\mbox{Sport}\xrightarrow{Players}\mbox{Athlete}\xrightarrow{leds}\mbox{Team}$&0.321&0.061                    \\

\bottomrule[1.5pt]
\end{tabular}}
\end{table}

\subsection{Meta-path Analysis}
We analyze and select high-quality meta-paths generated by \model, detailed in Table \ref{tab_example_metapath}. These meta-paths offer insightful interpretations and semantically accurate explanations across target relations for both general (e.g., \textit{Person}) and specific (e.g., \textit{Scientist}, \textit{Journalist}) entity types. 
Our model distinctively excels at leveraging synonymous relations to refine explanations or modify existing meta-paths with synonyms, demonstrating a strength inherent to LLMs.
\section{Ablation Studies}
\label{sec:ablation}
This section presents extensive ablation studies to assess the utility and sensitivity of our model's components. Specifically, we analyze the effects of varying meta-path sample prioritization, different prompt designs (with particular attention to the inclusion of meta-path samples), different LLM selections, removal of the atom selector, and removal of the meta-path cleaner.

\begin{table}[htb]
\label{tab_abl_priority}
\centering
\caption{Link prediction results (shown in ROC-AUC) on the YAGO26K-906 dataset over five runs using different replay strategies. Bold and underlined figures represent the best and second-best performances, respectively.}
\renewcommand\arraystretch{1}
\resizebox{0.65\linewidth}{!}{%
\begin{tabular}{cc|ccc}
\toprule[1.5pt]
&        & Conf. + Cov. & Conf. & Cov.  \\ \hline
\multirow{3}{*}{isCitizenOf}   & Direct & 0.936        & 0.925 & 0.911 \\
& Rank-based   & \textbf{0.949}        & 0.941 & \underline{0.944} \\
& Random & \multicolumn{3}{c}{0.876}   
\\ \hline
\multirow{3}{*}{DiedIn}        & Direct & 0.876        & 0.873 & 0.869 \\
& Rank-based   & \textbf{0.901}        & 0.885 & \underline{0.887} \\
& Random & \multicolumn{3}{c}{0.834}    \\ \hline
\multirow{3}{*}{GraduatedFrom} & Direct & \underline{0.828}        & 0.788 & 0.801 \\
& Rank-based   & \textbf{0.831}        & 0.812 & 0.82  \\
& Random & \multicolumn{3}{c}{0.768}  \\ 
\bottomrule[1.5pt]
\end{tabular}
}
\end{table}

\subsection{Analysis of Meta-path Sample Prioritization}
As outlined in Section~\ref{sec:framework}, our replay buffer could employ coverage, confidence, and their combination as priority and use either direct or rank-based prioritization, resulting in six meta-path sampling strategies. Besides, a random-sampling strategy is also applicable. Here, we empirically evaluate these seven strategies on the YAGO26K-906 dataset, as shown in Table~\ref{tab_abl_priority}. 
The results demonstrate that employing a combination of confidence and coverage scores for rank-based prioritization yields superior performance. Furthermore, all prioritization-based methods outperform random sampling, underscoring the importance of developing few-shot retrieval strategy specifically for meta-path discovery to enhance HIN reasoning.

\begin{figure}[h]
\centering
\includegraphics[width=1\linewidth]{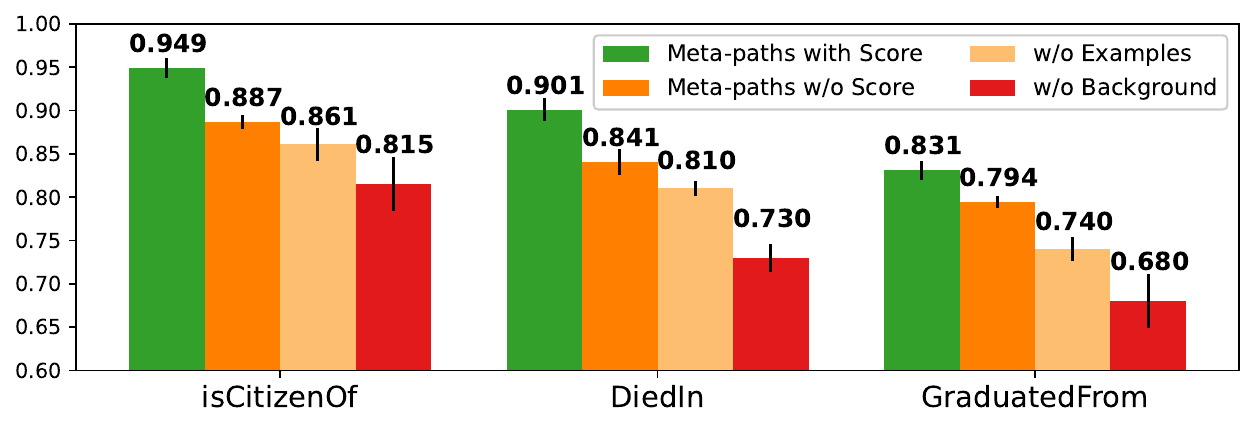}
\caption{Average link prediction performance (shown in ROC-AUC) across different prompt designs over five runs.}\label{fig_abl_prompt}
\end{figure} 
\begin{table}[htbp]
    \centering
    \caption{Meta-path Generated without Examples}
\resizebox{0.9\linewidth}{!}{
\begin{tabular}{c|c|c|c}
    \toprule[1.5pt]
\textbf{Relations}& \textbf{Meta-path}&\textbf{Conf.}&\textbf{Cov.}
\\ \midrule[1.1pt]
\multirow{5}{*}{\textit{isCitizenOf}}   
& $\mbox{Scientist} \xrightarrow{DiedIn}\mbox{Administrative}\xrightarrow{HasCaptial^{-1}}\mbox{Country}$   &0.132&0.020                     \\

& $ \mbox{University} \xrightarrow{GraduatedFrom^{-1}}\mbox{Person}\xrightarrow{LivesIn}\mbox{Country}$ &0.280 &0.005                       \\
             
& $\mbox{Writer}\xrightarrow{WasBornIn}\mbox{Commune}\xrightarrow{LocatedIn}\mbox{Country}$      &  0.091 & 0.005                   \\

&  $\mbox{Award} \xrightarrow{WonPrize^{-1}}\mbox{Scientist}\xrightarrow{LivesIn}\mbox{Country}$      &  0.070 & 0.041                   \\

&  $\mbox{OfficeHolder} \xrightarrow{AffiliatedTo}\mbox{Party}\xrightarrow{LocatedIn}\mbox{Country}$&0.056 &0.005     \\     
\hline
\multirow{5}{*}{\textit{WorksFor}}   

& $\mbox{Person} \xrightarrow{Collaborates}\mbox{Organization}$      &  0.235 & 0.008                   \\
& $\mbox{Writer} \xrightarrow{CotrolledBy}\mbox{Company}$          &0.259&0.002        \\
& $\mbox{Journalist} \xrightarrow{Collaborates}\mbox{Organization}$      &  0.333 & 0.001                   \\
&$\mbox{CEO} \xrightarrow{Colloborates}\mbox{Company}$ &0.344&0.004                 \\
&$\mbox{Person} \xrightarrow{HiredBy}\mbox{Store}\xrightarrow{Has}\mbox{Person}$      &  0.019 & 0.001                   \\
\bottomrule[1.5pt]
\end{tabular}}
\label{tab_without_ex}
\end{table}

\subsection{Analysis of Prompt Design}
We evaluate the impact on performance of different prompt designs by comparing the performance based on prompts with the score signal removed, without the few-shot meta-paths section, and lacking a background description.
Experiments are conducted on the YAGO26K-906 dataset, with results presented in Figure~\ref{fig_abl_prompt}.
Our findings indicate that incorporating an additional scoring signal significantly enhances the generation of high-quality meta-paths, with an average improvement of 6.26\% over scenarios without score signals. 
Furthermore, supplying meta-path samples yields, an average 4.71\% improvement compared to not providing any few-shot examples. 
Most notably, we discovered that providing a background description is crucial; without it, LLMs fail to generate valid meta-paths, resulting in poor outcomes.

Besides, to assess if providing meta-path demonstration samples mitigate knowledge conflicts, we compared meta-paths generated without them (Table 7) to those produced by our model (Table 5). Meta-paths generated without demonstration samples generally show lower confidence and coverage scores. As illustrated in Table 7, while some meta-paths align with commonsense, the majority of the generated meta-paths demonstrate lower confidence and coverage, reflecting inconsistencies with the knowledge embedded in the HIN. This observation explicitly validates that our methods effectively resolve knowledge conflicts.

\begin{table}[htb]
\centering
\caption{Performance for link prediction (shown in ROC-AUC) with different LLMs on the YAGO26K-906 dataset, where bold and underlined values denote the best and second-best results, respectively.
} \renewcommand\arraystretch{1.1}
\resizebox{0.75\linewidth}{!}{%
\begin{tabular}{cccc}
\toprule[1.5pt]
& isCitizenOf & DiedIn & GraduatedFrom \\ \hline 
GPT-4               & 0.925      & 0.877 & \textbf{0.846}         \\
ChatGLM             & 0.906       & 0.883 & 0.800        \\
Mistral-7B-Instruct & 0.928      & 0.872 & 0.795        \\
Llama2-chat-7B      & \textbf{0.949}       & \textbf{0.901}  & \underline{0.831}         \\
Llama2-chat-13B     & \underline{0.929}       & \underline{0.892} & 0.787        \\
Llama2-chat-70B     & 0.906      & 0.865  & 0.801   
\\
\bottomrule[1.5pt]
\end{tabular}
}
\label{tab_abl_llm}
\end{table}
\subsection{Analysis of LLM Selection}
The foundational aspect of our work, the proficiency of LLMs in understanding natural languages, led us to evaluate \model~across various LLMs, including GPT-4~\citep{openai2023gpt}, ChatGLM~\citep{zeng2022glm}, Mistral-7B-Instruct~\citep{jiang2023mistral}, and LLaMA2-Chat~\citep{touvron2023llama} at scales of 7B, 13B, and 70B.
Results in Table~\ref{tab_abl_llm} demonstrate that \model~achieves relatively consistent performance across different LLMs, underscoring its robustness to LLM selection.
Moreover, using larger LLMs does not guarantee improved performance, potentially due to a higher probability of early termination from increased computational costs, leading to a reduced number of rules that slows the rule evolution process. This observation suggests the preference for a compact LLM model with high output efficiency in our framework.

\subsection{Analysis of Atom Selector}
\begin{table}[tb]
\centering
\caption{Performance for link prediction (shown in ROC-AUC) and efficiency with and without atom selector on the NELL dataset.
}
\renewcommand\arraystretch{1}
\resizebox{0.75\linewidth}{!}{%
\begin{tabular}{c|cc|cc|cc}
\toprule[1.5pt]
& \multicolumn{2}{c|}{AUC} & \multicolumn{2}{c|}{Time / Round (s)} & \multicolumn{2}{c}{Error Rate (\%)} \\ 
& with       & w/o        & with               & w/o                 & with              & w/o             \\ \hline
WorkFor      & 0.897      & 0.816      & 43.5               & 100.2               & 6.4               & 21.7             \\
PlaysAgainst & 0.957      & 0.879      & 39.5               & 99.3                & 7.8               & 24.4             \\
CompetesWith & 0.773      & 0.732      & 33.4               & 95.7                & 7.6               & 27.1            
\\
\bottomrule[1.5pt]
\end{tabular}
}
\label{tab_ablation_atom}
\end{table}
To assess the atom selector's effectiveness in accurately selecting taxonomies for meta-path discovery and its potential to reduce time, we conducted link prediction tests on the complex HIN of NELL. The results in Table~\ref{tab_ablation_atom} indicate that without the atom selector, performance decreases significantly by 7.49\% and processing speed slows by approximately 2.5 times. Furthermore, the error rate escalates by a factor of 3.36, which is attributable to the challenges of processing overlength inputs that impair the LLM's comprehension and the complexity of combining a vast taxonomy.

\subsection{Analysis of Meta-path Cleaner}
\begin{table}[tb]
\centering
\caption{Performance for link prediction (shown in ROC-AUC) and efficiency with and without meta-path cleaner on the NELL dataset.
}
\renewcommand\arraystretch{1.35}
\resizebox{0.6\linewidth}{!}{%
\begin{tabular}{c|ccc}
\toprule[1.5pt]
& WorkFor & PlaysAgainst & CompetesWith\\ \hline
with  cleaner     & 0.897      & 0.957      & 0.773      \\
w/o cleaner & 0.863      & 0.932      & 0.761  
\\
\bottomrule[1.5pt]
\end{tabular}
}
\label{tab_ablation_cleaner}
\end{table}
The metapath cleaner, the final remedial module, identifies and corrects errors in generated meta-paths. To evaluate its impact, we conducted link prediction tests on the complex HIN of NELL, excluding the cleaner. Table~\ref{tab_ablation_atom} shows a performance drop of 2.7\% without the Meta-path Cleaner. Although this drop is less significant compared to removing other modules or designs, it underscores the importance of the cleaner in generating and preserving high-quality meta-paths.

\section{Conclusion}
In this paper, we synergize the natural language processing proficiency of LLMs with meta-path discovery in HINs. Through carefully crafted components that address corpora bias, lexical discrepancies, and hallucination, our model adeptly leverages LLMs to generate high-quality meta-paths. 
Corpora bias is addressed through in-context learning by using sampled meta-paths (with proritization) from HIN to guide generation. Lexical discrepancy is managed with designed prompts and constraining the LLM to candidate taxonomy. To reduce hallucination, we use positive examples in in-context learning, supplemented with background information and specific constraints to prevent generating unlawful rules.

The generated meta-paths are utilized in both transductive and inductive reasoning tasks within HINs, achieving notable results. Ablation studies highlight the importance of each model component.
Our findings demonstrate that the commonsense knowledge embedded in LLMs can be translated into explicit meta-paths. By using in-context prompting techniques, we eliminate knowledge conflicts and generate reliable meta-paths for offline HINs. 

Our model is capable of generating high-quality meta-paths efficiently, offering significant advantages in various real-world HIN applications requiring rapid, reliable, and explainable responses. This capability is particularly useful in commercial activities and drug-target interaction prediction. 
The generated meta-paths could identify related, meaningful instance paths, facilitating reasoning and providing essential information for accurate predictions for queried entities. 
Further, such generation-retrieval-prediction process can achieve seamless three-step integration with LLMs, making it more automated, trustworthy, and traceable compared to current reasoning approaches using black-box LLMs.

Our future research will focus on enhancing LLMs' understanding of knowledge embedded in HINs and integrating a chain-of-thought mechanism into the discovery process. We will also explore leveraging LLMs to produce rules for reasoning or predicting across other types of offline databases.

\appendix
\section{Example Prompt}
The following prompt is used to generate meta-paths for the relation "isCitizenOf":

\label{sec:appendix_prompt}
\textit{
Background: "Within Heterogeneous Information Networks (HINs), a meta-path represents a defined sequence of relations among multiple entity types in the network. Each meta-path should start and end with an entity type, involving a series of interactions between types and relations"}

\textit{Few-shot Example: "Here are some example meta-paths and their scores for \{isCitizenOf\}: ..." }

\textit{Requirement: "Please generate as many meta-paths as possible to explain relation  \{isCitizenOf\}. You need to generate meta-paths with \{$L*2+1$\} words in total. Relations and types in the meta-paths must be selected from \{relations\} and \{types\} separately. Do not return any explanation."
}

\section{Baseline Description}
\label{sec:appendix_baseline}
\noindent 
\begin{itemize}
    \item \textbf{MPDRL~\citep{wan2020mpdrl}.} MPDRL employs an RL agent to identify path instances and summarizes them as meta-paths.

    \item \textbf{PCRW~\citep{lao2010pcrw}.} PCRW discovers path instances through random walks before summarization.

    \item \textbf{Autopath~\citep{yang2018autopath}.} Autopath computes the similarity between entity pairs using empirical arrival probabilities at the tail entity. We optimized parameters for the best results.

    \item \textbf{Metapath2Vec~\citep{dong2017metapath2vec}.} Metapath2Vec constructs node-level embeddings via meta-path-based random walks.

    \item \textbf{HIN2Vec~\citep{fu2017hin2vec}.} We used code from the authors and optimized parameters.

    \item \textbf{HGMAE~\citep{tian2023heterogeneous}.} HGMAE is a masked auto-encoder model that leverages meta-path masking and adaptive attribute masking with a dynamic mask, facilitating effective and stable learning on complex graph structures.

    \item \textbf{RotatE~\citep{sun2019rotate}.} RotatE learns embeddings to represent entities and relations in KBs, by modeling relation as rotations in the whole complex space. We adopt the best parameters reported in their paper for Yago and apply the same for other datasets.

    \item \textbf{TransE~\citep{bordes2013transe}.} 
TransE models embeddings by aligning the sum of head and relation vectors closely with the tail vector.

    \item \textbf{DistMult~\citep{yang2014Distmult}.} DistMult identifies entity relationships through the Hadamard product of embeddings.

    \item \textbf{ComplEx~\citep{trouillon2016complex}.} ComplEx extends DistMult into the complex space, achieving similar performance.

    \item \textbf{RNNLogic~\citep{qu2020rnnlogic}.} RNNLogic integrates an RNN with a logic reasoning module for rule-learning and knowledge encoding.

    \item \textbf{MINERVA~\citep{das2018minerva}.} MINERVA employs a neural RL-based multi-hop method. In our link prediction experiments, we enhance results by computing predicted scores for both positive and negative samples across entity pairs and deriving similarity through a Softmax operation on these scores, mirroring their NELL evaluation technique but surpassing it in effectiveness.

    \item \textbf{MLN4KB~\citep{fang2023mln4kb}.} MLN4KB is an efficient reasoning method based on Markov logic network.
\end{itemize}

\section{Dataset}
\label{sec:appendix_data}
\begin{itemize}
    \item \textbf{YAGO26K-906~\citep{suchanek2007yago,hao2019joie}.} YAGO is a knowledge base (KB) curated from Wikipedia and WordNet facts. The initial version of YAGO had limited semantic relations between entity types and we use a version refined by Hao et al., which enriches the core YAGO facts with an expanded taxonomy.

    \item \textbf{Dbpedia~\citep{auer2007dbpedia}.} Dbpedia is a comprehensive KB extracted from Wikipedia spanning numerous specific domains and general knowledge areas. The reasoned relations are: \textit{musicalBand, musicalArtist, subsequentWork, nationality, spouse, countySeat, stateOfOrigin, distributingCompany, distributingLabel, parent, trainer}.

    \item \textbf{NELL~\citep{mitchell2018never}.} NELL is a KB generated from over 500 million unstructured web pages. We work with the preprocessed portion numbered 1115.
\end{itemize}

\section*{Data availability}
We have shared the link to the code at the Attach File step. The link is: \url{https://data.mendeley.com/preview/bhtwbs92fy?a=bf23a210-76e1-474b-8d61-210ef59b6bc1}

\printcredits

\bibliographystyle{apalike} 
\bibliography{casrefs}

\vskip3pt
\clearpage
\bio{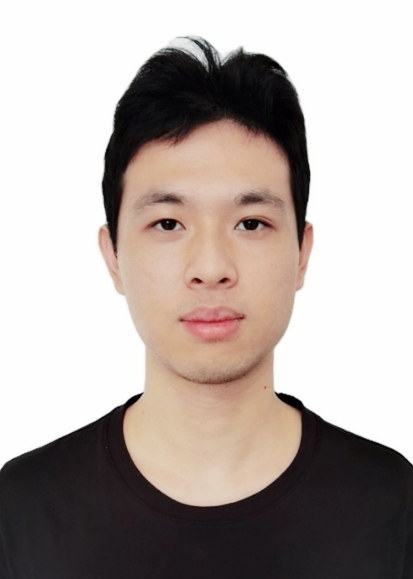}
{\bf Shixuan Liu}\
received his B.S. and Ph.D. degrees from the National University of Defense Technology, Changsha, China, in 2019 and 2024, respectively. He is also a visiting scholar in the Department of Computer Science and Technology at Tsinghua University, where he has spent two years. He has published over 10 papers in prestigious journals and conferences, including T-PAMI, T-KDE, T-CYB, and ICDM, focusing on knowledge reasoning and data mining.
\endbio
\vspace{60pt}

\bio{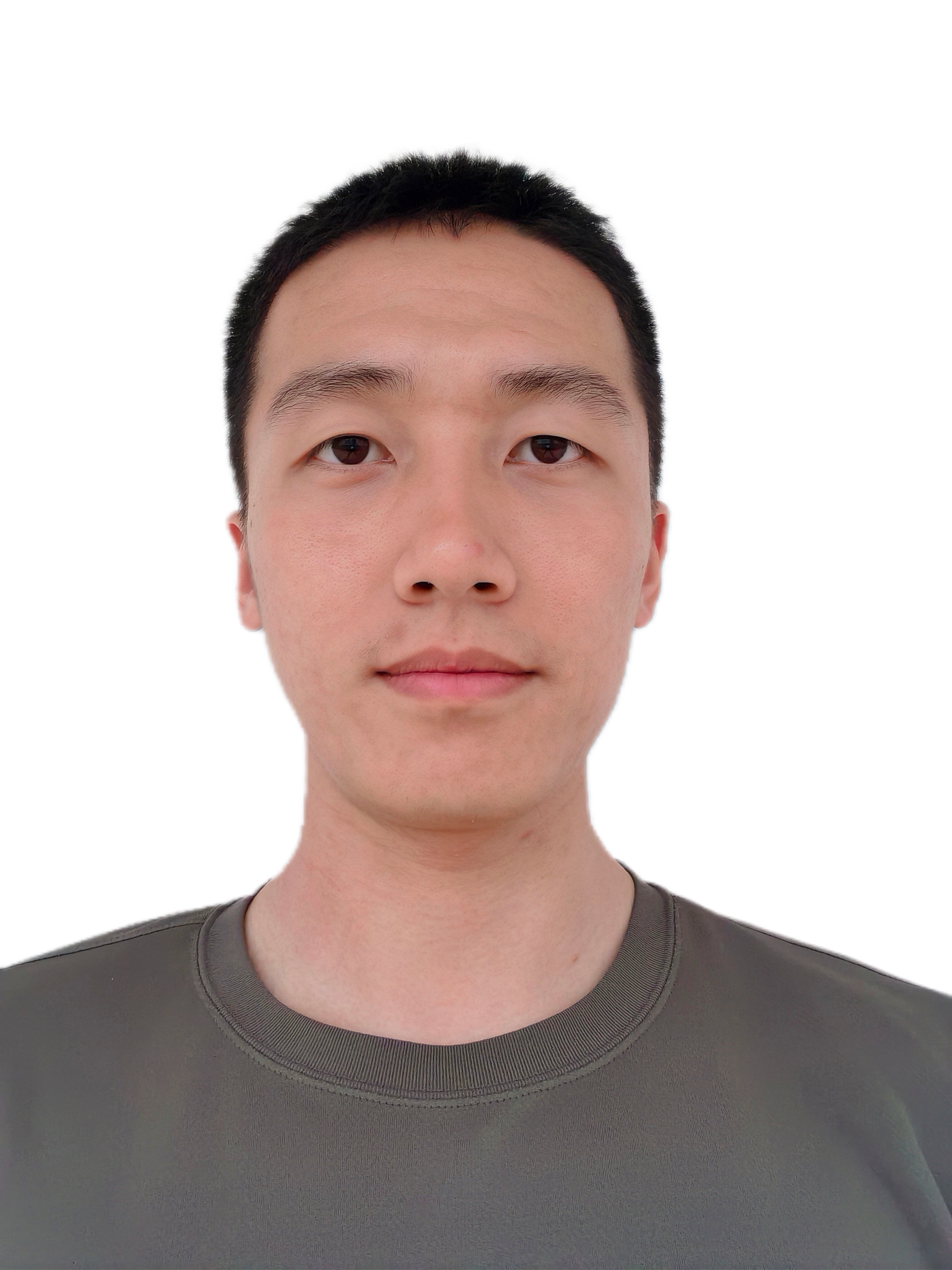}
{\bf Haoxiang Cheng}\
received the B.S. degree in systems engineering from the National University of Defense Technology, Changsha, China, in 2024, where he is currently pursuing the master's degree. His research interests include Large Language Models, and knowledge reasoning.
\endbio

\vspace{60pt}
\bio{bio_Y_Wang.jpg}
{\bf Yunfei Wang}\
received the B.S. degree in civil
engineering from the Hunan University, Changsha, China, in 2020. She is now pursuing the Ph.D degree at the National University of Defense Technology, Changsha, China. Her research interests include auto penetration test, reinforcement learning and cyber-security.
\endbio

\vspace{60pt}
\bio{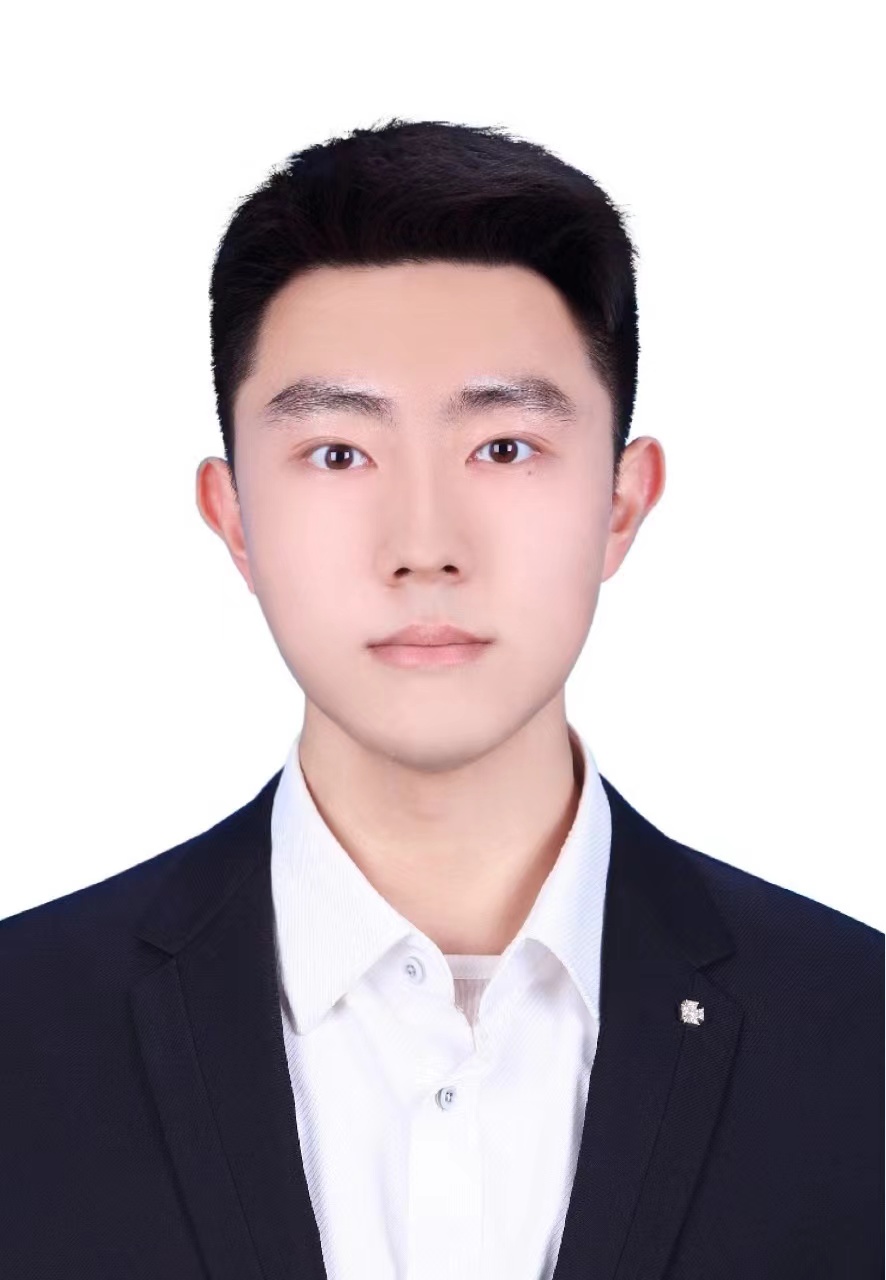}
{\bf Yue He}\
is a Postdoctoral fellow at Tsinghua University. He received his Ph.D. in the Department of Computer Science and Technology from Tsinghua University in 2023. His research interests include out-of-distribution generalization, causal structure learning, and graph computing. He has published more than 20 papers in prestigious conferences and journals in machine learning, data mining, and computer vision. He serves as a PC member in many academic conferences, including ICML2024, Neurips2024, UAI2024 and etc.
\endbio

\vspace{60pt}
\bio{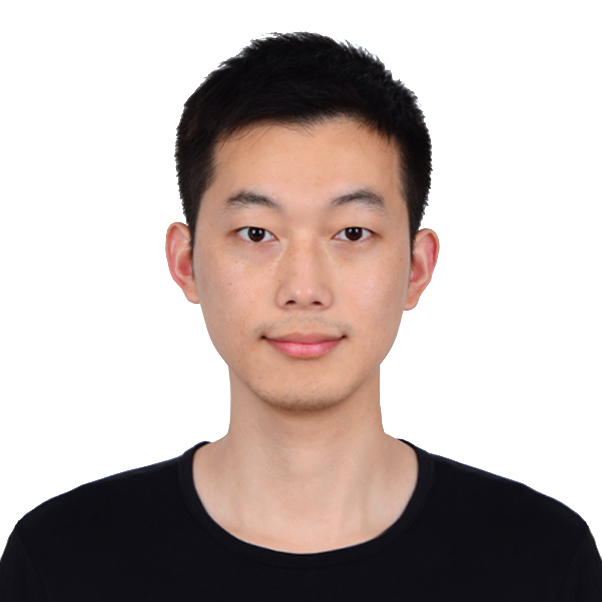}
{\bf Changjun Fan}\
received the B.S. degree, M.S. degree and PhD degree all from National University of Defense Technology, Changsha, China, in 2013, 2015 and 2020. He is also a visiting scholar at Department of Computer Science, University of California, Los Angeles, for two years.  He is currently an associate professor at National University of Defense Technology, China. His research interests include deep graph learning and complex systems, with a special focus on their applications on intelligent decision making. During his previous study, he has published a number of refereed journals and conference proceedings, such as Nature Machine Intelligence, Nature Communications, AAAI, CIKM, etc.
\endbio

\bio{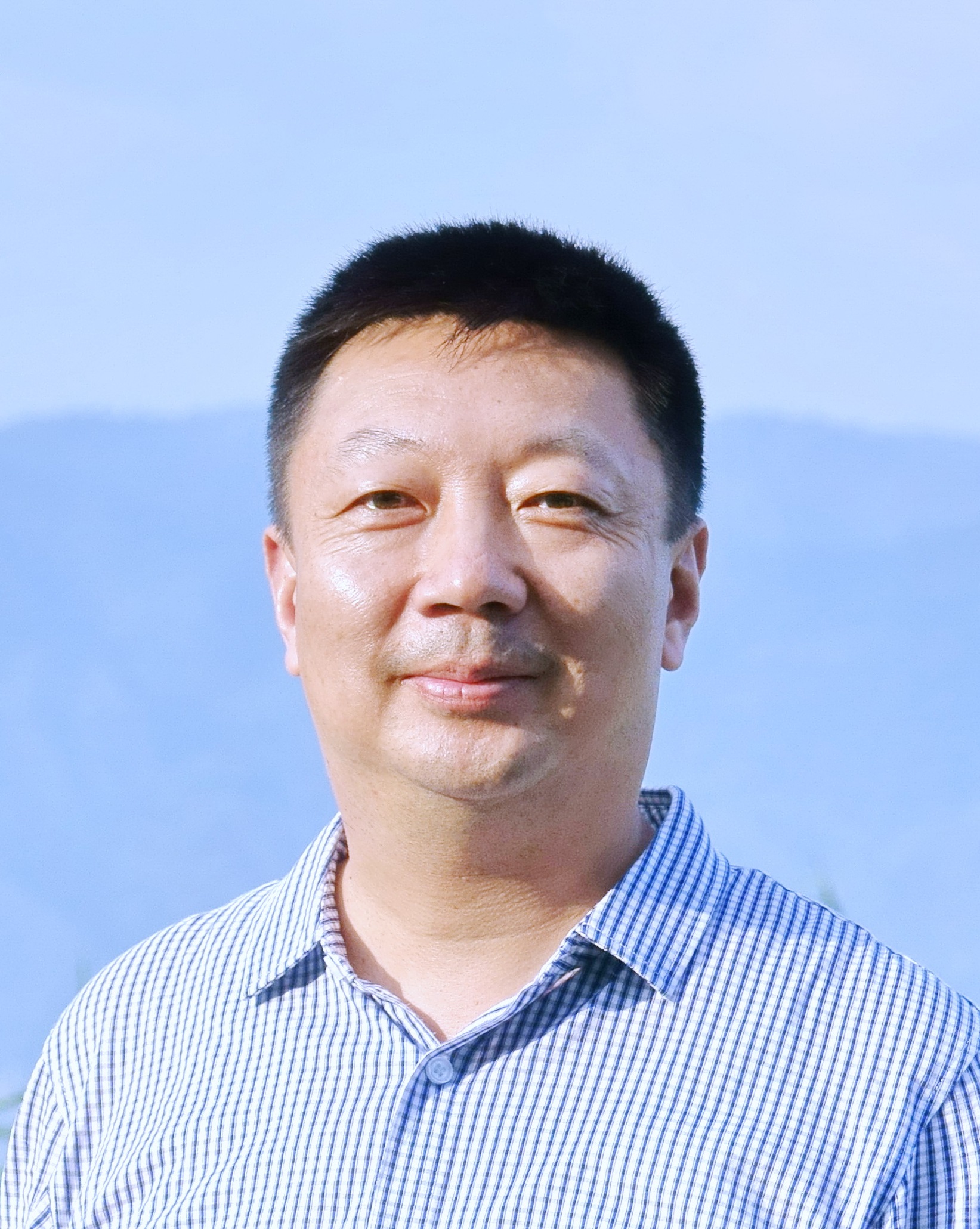}
{\bf Zhong Liu}\
received the B.S. degree in Physics from Central China Normal University, Wuhan, Hubei, China, in 1990, the M.S. degree in computer software and the Ph.D. degree in management science and engineering both from National University of Defense Technology, Changsha, China, in 1997 and 2000. He is a professor in the College of Systems Engineering, National University of Defense Technology, Changsha, China. His research interests include intelligent information systems, and intelligent decision making.
\endbio

\end{document}